\documentclass[12]{iopart}

\usepackage{graphicx}
\usepackage{bm}

\begin{document}

\pagestyle{empty}
\begin{center}
{\Large \bfseries
Energy dependence of fluctuations in central Pb+Pb collisions
from NA49 at the CERN SPS}
\end{center}
\vspace*{0.2cm}
\renewcommand{\thefootnote}{\fnsymbol{footnote}}
\noindent

M. Rybczy\'nski$^{11,\S}$\footnote[0]{\S~talk presented at Quark Matter 2008},C.~Alt$^{9}$, T.~Anticic$^{23}$, B.~Baatar$^{8}$,D.~Barna$^{4}$,
J.~Bartke$^{6}$, L.~Betev$^{10}$, H.~Bia{\l}\-kowska$^{20}$,
C.~Blume$^{9}$,  B.~Boimska$^{20}$, M.~Botje$^{1}$,
J.~Bracinik$^{3}$, R.~Bramm$^{9}$, P.~Bun\v{c}i\'{c}$^{10}$,
V.~Cerny$^{3}$, P.~Christakoglou$^{2}$, P.~Chung$^{19}$, O.~Chvala$^{14}$,
J.G.~Cramer$^{16}$, P.~Csat\'{o}$^{4}$, P.~Dinkelaker$^{9}$, V.~Eckardt$^{13}$,
D.~Flierl$^{9}$, Z.~Fodor$^{4}$, P.~Foka$^{7}$, V.~Friese$^{7}$, J.~G\'{a}l$^{4}$,
M.~Ga\'zdzicki$^{9,11}$, V.~Genchev$^{18}$, E.~G{\l}adysz$^{6}$, K.~Grebieszkow$^{22}$,
S.~Hegyi$^{4}$, C.~H\"{o}hne$^{7}$, K.~Kadija$^{23}$, A.~Karev$^{13}$, D.~Kikola$^{22}$,
M.~Kliemant$^{9}$, S.~Kniege$^{9}$, V.I.~Kolesnikov$^{8}$, E.~Kornas$^{6}$,
M.~Kowalski$^{6}$, I.~Kraus$^{7}$, M.~Kreps$^{3}$, A.~Laszlo$^{4}$, R.~Lacey$^{19}$,
M.~van~Leeuwen$^{1}$, P.~L\'{e}vai$^{4}$, L.~Litov$^{17}$, B.~Lungwitz$^{9}$,
M.~Makariev$^{17}$, A.I.~Malakhov$^{8}$, M.~Mateev$^{17}$, G.L.~Melkumov$^{8}$,
A.~Mischke$^{1}$, M.~Mitrovski$^{9}$, J.~Moln\'{a}r$^{4}$, St.~Mr\'owczy\'nski$^{11}$,
V.~Nicolic$^{23}$, G.~P\'{a}lla$^{4}$, A.D.~Panagiotou$^{2}$, D.~Panayotov$^{17}$,
A.~Petridis$^{2,\dagger}$, W.~Peryt$^{22}$, M.~Pikna$^{3}$, J.~Pluta$^{22}$, D.~Prindle$^{16}$,
F.~P\"{u}hlhofer$^{12}$, R.~Renfordt$^{9}$, C.~Roland$^{5}$, G.~Roland$^{5}$,
A.~Rybicki$^{6}$, A.~Sandoval$^{7}$, N.~Schmitz$^{13}$, T.~Schuster$^{9}$, P.~Seyboth$^{13}$,
F.~Sikl\'{e}r$^{4}$, B.~Sitar$^{3}$, E.~Skrzypczak$^{21}$, M.~Slodkowski$^{22}$,
G.~Stefanek$^{11}$, R.~Stock$^{9}$, C.~Strabel$^{9}$, H.~Str\"{o}bele$^{9}$, T.~Susa$^{23}$,
I.~Szentp\'{e}tery$^{4}$, J.~Sziklai$^{4}$, M.~Szuba$^{22}$, P.~Szymanski$^{10,20}$,
V.~Trubnikov$^{20}$, M.~Utvi\'{c}$^{9}$, D.~Varga$^{4,10}$, M.~Vassiliou$^{2}$,
G.I.~Veres$^{4,5}$, G.~Vesztergombi$^{4}$, D.~Vrani\'{c}$^{7}$, A.~Wetzler$^{9}$,
Z.~W{\l}odarczyk$^{11}$, A.~Wojtaszek$^{11}$, I.K.~Yoo$^{15}$, J.~Zim\'{a}nyi$^{4,\dagger}$
\begin{center}
(NA49 Collaboration)
\end{center}

\vspace{0.5cm}
\noindent
$^{1}$NIKHEF, Amsterdam, Netherlands. \\
$^{2}$Department of Physics, University of Athens, Athens, Greece.\\
$^{3}$Comenius University, Bratislava, Slovakia.\\
$^{4}$KFKI Research Institute for Particle and Nuclear Physics, Budapest, Hungary.\\
$^{5}$MIT, Cambridge, USA.\\
$^{6}$Henryk Niewodniczanski Institute of Nuclear Physics, Polish Academy of Sciences, Cracow, Poland.\\
$^{7}$Gesellschaft f\"{u}r Schwerionenforschung (GSI), Darmstadt, Germany.\\
$^{8}$Joint Institute for Nuclear Research, Dubna, Russia.\\
$^{9}$Fachbereich Physik der Universit\"{a}t, Frankfurt, Germany.\\
$^{10}$CERN, Geneva, Switzerland.\\
$^{11}$Institute of Physics, Jan Kochanowski University, Kielce, Poland.\\
$^{12}$Fachbereich Physik der Universit\"{a}t, Marburg, Germany.\\
$^{13}$Max-Planck-Institut f\"{u}r Physik, Munich, Germany.\\
$^{14}$Charles University, Faculty of Mathematics and Physics, Institute of Particle and Nuclear Physics, Prague, Czech Republic.\\
$^{15}$Department of Physics, Pusan National University, Pusan, Republic of Korea.\\
$^{16}$Nuclear Physics Laboratory, University of Washington, Seattle, WA, USA.\\
$^{17}$Atomic Physics Department, Sofia University St. Kliment Ohridski, Sofia, Bulgaria.\\
$^{18}$Institute for Nuclear Research and Nuclear Energy, Sofia, Bulgaria.\\
$^{19}$Department of Chemistry, Stony Brook Univ. (SUNYSB), Stony Brook, USA.\\
$^{20}$Institute for Nuclear Studies, Warsaw, Poland.\\
$^{21}$Institute for Experimental Physics, University of Warsaw, Warsaw, Poland.\\
$^{22}$Faculty of Physics, Warsaw University of Technology, Warsaw, Poland.\\
$^{23}$Rudjer Boskovic Institute, Zagreb, Croatia.\\
$^{\dagger}$deceased

\begin{center}
\begin{parbox}[t]{12cm}{
{\small
{\bfseries Abstract:} The latest NA49 results on fluctuations of
multiplicity and average transverse-momentum analyzed on an event-by-event basis are presented for central Pb+Pb
interactions over the whole SPS energy range (20A - 158A GeV).
The scaled variance of the multiplicity distribution decreases with collision energy
whereas the $\Phi_{p_{T}}$ measure of $\langle p_T \rangle$
fluctuations is small and independent of collision energy.
Thus in central Pb+Pb collisions these fluctuations do
not show an indication of the critical point of
strongly interacting matter.
}}
\end{parbox}
\end{center}

\setlength{\textwidth}{11cm}
\section{Introduction}

High-energy nucleus-nucleus collisions have been studied over the last two decades.
The main goal of these efforts is to understand the properties of strongly
interacting matter under extreme conditions of very high energy where the creation of the quark-gluon plasma
(QGP) is expected~\cite{Collins,Shuryak}. Various
collision characteristics and their energy dependence suggest that
a transient state of deconfined matter may be created at
collision energies as low as $30\,A$~GeV~\cite{Gazdzicki}.
Fluctuations in physical observables in heavy-ion collisions have been
a topic of particular interest in recent years as they may provide
important signals regarding the formation of QGP and the existence of a critical point.

The NA49 detector~\cite{A.1}, located at the CERN SPS, offers the
unique possibility to study the energy dependence of various
collision characteristics on an event-by-event basis. The NA49
energy-scan program obtained results on several kinds of
event-by-event fluctuations in central Pb+Pb collision. For
example, the observed net-charge fluctuations~\cite{net-charge}
are close to zero and much larger than expected for an ideal gas
of deconfined quarks and gluons. On the other hand the dynamical
$K/\pi$ fluctuations~\cite{kpi} show an increase at low SPS
energies, which is qualitatively consistent with the expectation
for the onset of deconfinement~\cite{Gorenstein:2003hk}. Recently
it was suggested in~\cite{stephanov} that non-monotonic energy
dependence of multiplicity and transverse-momentum fluctuations
can help to locate the second-order critical end-point. Therefore,
the energy-scan program analyzed the event-by-event fluctuations
of these observables. In this report the energy dependence of
multiplicity and transverse-momentum fluctuations in very central
Pb+Pb collisions measured by the NA49 experiment is presented and
compared to predictions of the various hadron-gas and
string-hadronic models.

\section{Experiment, Data and Analysis}

NA49~\cite{A.1} is a fixed target experiment located at the CERN
SPS which comprises four large-volume Time Projection Chambers
(TPC). Two chambers, the Vertex TPCs (VTPC-1 and VTPC-2), are
located in the magnetic field of two superconducting dipole
magnets ($1.5$ and $1.1\,\mathrm{T}$), while the two others
(MTPC-L and MTPC-R) are positioned downstream of the magnets
symmetrically to the beam line. A zero degree calorimeter at the
downstream end of the experiment is used to trigger on or select the
centrality of the collisions. The data presented here correspond
to the most central (1\% for the analysis of
multiplicity and 7.2\% for the analysis of
transverse-momentum fluctuations) events for 20A, 30A, 40A, 80A
and 158A GeV. The acceptance was restricted in the analysis
to the forward hemisphere:
$1<y_{\pi}<y_{beam}$ in the case of multiplicity fluctuations and
$1.1 < y_{\pi} < 2.6$$^*$~\footnote[0]{* $y_{\pi}$ is the particle
rapidity calculated in the center-of-mass reference system,
assuming the pion mass for particles.} in the case of transverse
momentum fluctuations. Note, that the full acceptance of the
NA49 detector changes with collision energy.

The scaled variance, $\omega$, used in this paper as a measure of
multiplicity fluctuations, is defined as
$\omega=\frac{Var(n)}{<n>}=\frac{<n^2>-<n>^2}{<n>}$ where
$Var(n)=\sum_n (n-\langle n\rangle)^2 P(n)$ and $\langle
n\rangle=\sum n \cdot P(n)$ are variance and mean of the
multiplicity distribution, respectively. Note, that in statistical
models~\cite{Begun:2006uu} $\omega$ is independent of volume (for
large enough volumes). Conservation laws decrease the value of the
scaled variance.

In the NA49 experiment the $\Phi_{p_{T}}$ fluctuation/correlation measure, proposed in~\cite{Gaz92},
is used to determine transverse-momentum fluctuations.
For a complete definition of $\Phi_{p_{T}}$ see~\cite{Gaz92}
and the publication of NA49~\cite{fluct_size}. $\Phi_{p_{T}}$
quantifies the difference between event-by-event fluctuations of
transverse momentum in data and the corresponding fluctuations in
'mixed' events. In the latter correlations vanish by construction. There are two important properties of the $\Phi_{p_{T}}$ measure.
When the system consists of independently emitted particles (no inter-particle correlations)
$\Phi_{p_{T}}$ assumes a value of zero. On the other hand, if A+A
collisions can be treated as an incoherent superposition of independent
N+N interactions (superposition model), then $\Phi_{p_{T}}$ has a
constant value, the same for A+A and N+N interactions.

\section{Search for the Critical Point}

In this section we show results on the energy dependence of multiplicity and transverse-momentum fluctuations, two observables which
may be sensitive to the presence of a critical end-point~\cite{stephanov}.

\subsection{Multiplicity Fluctuations}

\begin{figure}[b]

\begin{minipage}{7cm}
\includegraphics[width=.99\textwidth]{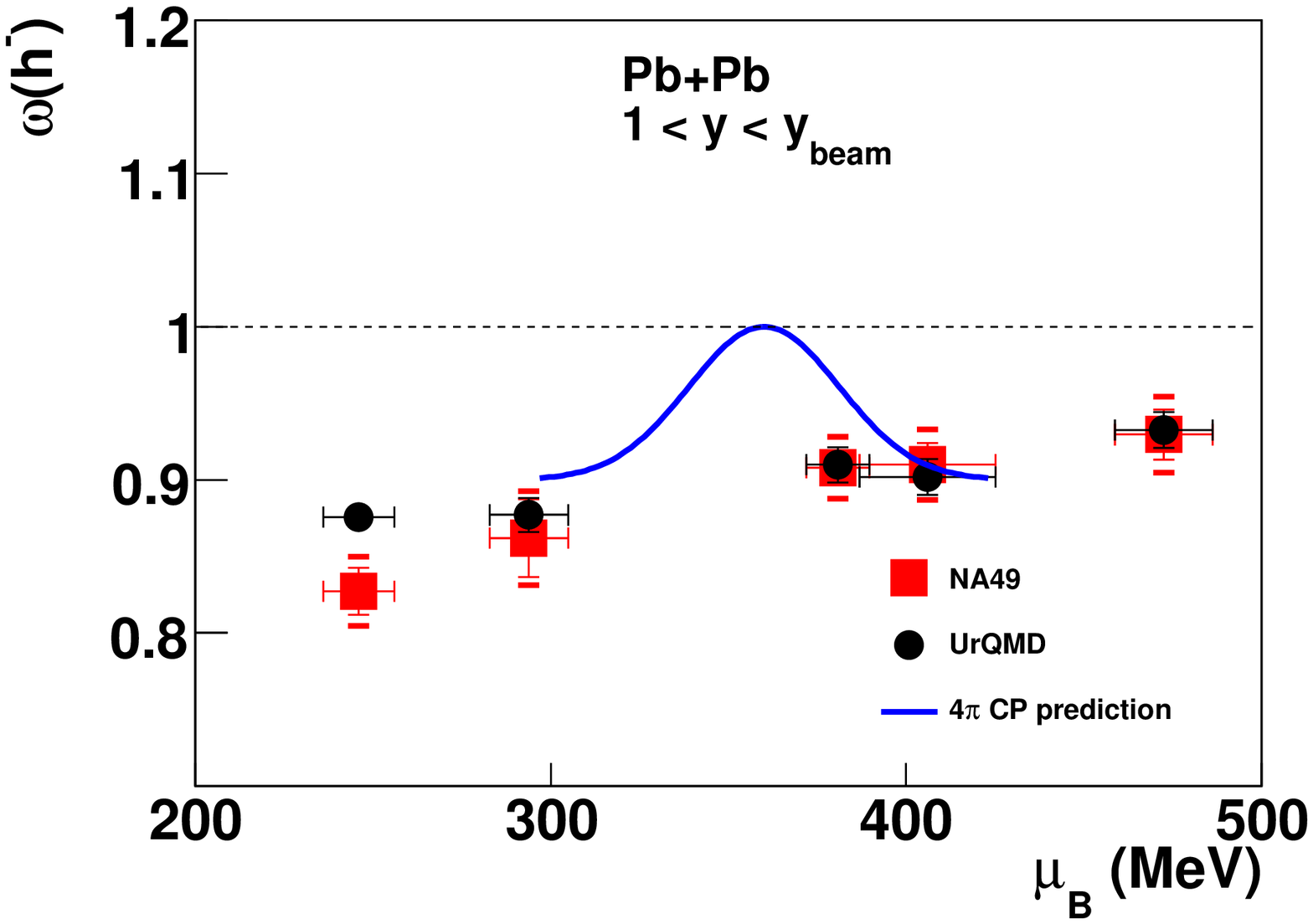}

\end{minipage}
\begin{minipage}{7cm}
\includegraphics[width=1.025\textwidth]{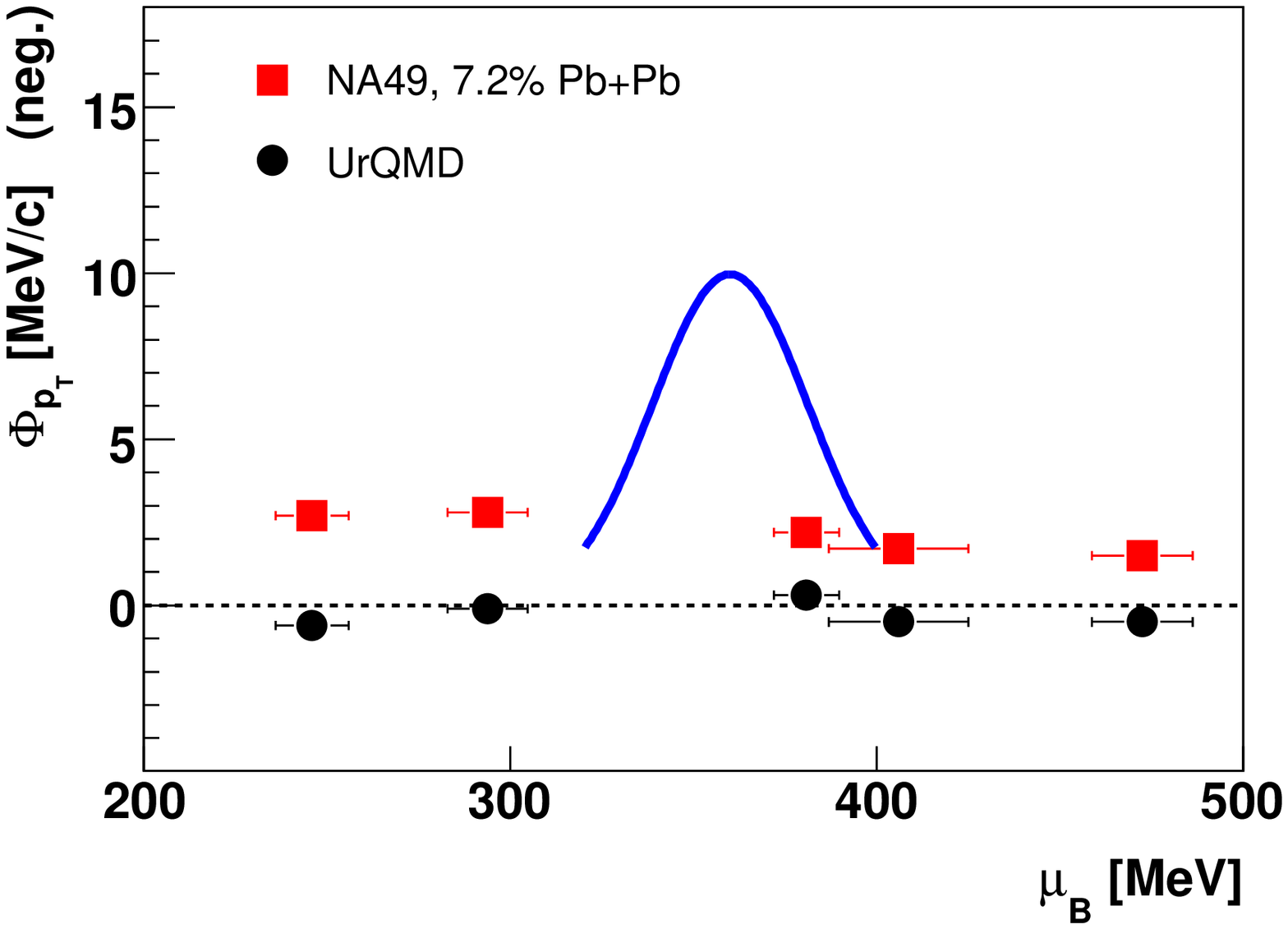}
\end{minipage}

\begin{center}
\begin{minipage}{11.8cm}
\begin{center}
\parbox[t]{5.2cm}{
\footnotesize{{\bf Figure 1:} (Color online) Scaled variance
$\omega(h^-)$ of the multiplicity distribution of negatively
charged hadrons as a function of baryo-chemical potential
$\mu_{B}$ for the NA49 data (squares) and the UrQMD~\cite{UrQMD}
model (circles). The curve shows the predicted values of the
scaled variance in $4\pi$ acceptance in the vicinity of the
critical end-point~\cite{fodor-katz,hatta-ikeda,stephanov}. The
effect of finite acceptance needs to be estimated. See text for
details. \label{fig:mult} }}\hfill
\parbox[t]{5.2cm}{
\footnotesize{{\bf Figure 2:} (Color online) $\Phi_{p_{T}}$ measure of transverse-momentum fluctuations calculated for negatively charged hadrons with transverse momenta $p_{T}\leq 500$~MeV/c as a function of baryo-chemical potential $\mu_{B}$ for the NA49 data (squares) and the UrQMD~\cite{UrQMD} model (circles). The curve shows the predicted $\Phi_{p_{T}}$ values in the vicinity of the critical end-point~\cite{stephanov}. See text for details. \label{fig:phipt}
}}
\end{center}
\end{minipage}
\end{center}
\end{figure}
Results on multiplicity fluctuations for negatively charged
hadrons are presented for Pb+Pb collisions at 20A, 30A, 40A, 80A
and 158A GeV. The measurements are plotted versus the
baryo-chemical potential $\mu_{B}$ which is derived from
hadron-gas model fits to the particle yields obtained by
NA49~\cite{hadron-gas}. In order to minimize the fluctuations in
the number of participants, the $1\%$ most central collisions
according to the energy of projectile spectators measured in the
veto calorimeter are selected. In figure~1 the scaled variance
$\omega$(h$^-$) of the multiplicity distribution of negatively
charged particles is shown as a function of baryo-chemical
potential $\mu_{B}$ and compared to the predictions of the
UrQMD~\cite{UrQMD} model. We also show the predicted peak of the
scaled variance in $4\pi$ acceptance in the vicinity of the
critical end-point~\cite{fodor-katz,hatta-ikeda,stephanov}. It is
evident from the plot that there is no indication of the critical
end-point in the energy dependence of multiplicity fluctuations.

\subsection{Transverse-momentum Fluctuations}

Following the suggestion~\cite{stephanov} that fluctuations due to the critical point should be dominated by fluctuations of pions
with $p_{T}\leq 500$~MeV/c  we analysed transverse momentum fluctuations for this restricted $p_T$ range.
Figure~2 shows $\Phi_{p_{T}}$ values for negatively, positively and all charged particles calculated from NA49 data
as a function of baryo-chemical potential $\mu_{B}$.
As in the case of multiplicity fluctuations we have also plotted the prediction of the values of $\Phi_{p_{T}}$ near the critical point$^*$\footnote[0]{* Note that predicted values of $\Phi_{p_{T}}$ at the critical point should result in $\Phi_{p_{T}}\simeq 20$~MeV/c, however the effect of limited acceptance of NA49 reduces them to $\Phi_{p_{T}}\simeq 10$~MeV/c}. Also for transverse-momentum fluctuations there is no significant energy dependence of the $\Phi_{p_{T}}$ measure when low transverse momenta are selected and thus there is no evidence for a critical point.

\section{Comparison to Models}

\begin{figure}[b]

\begin{minipage}{7cm}
\includegraphics[width=.99\textwidth]{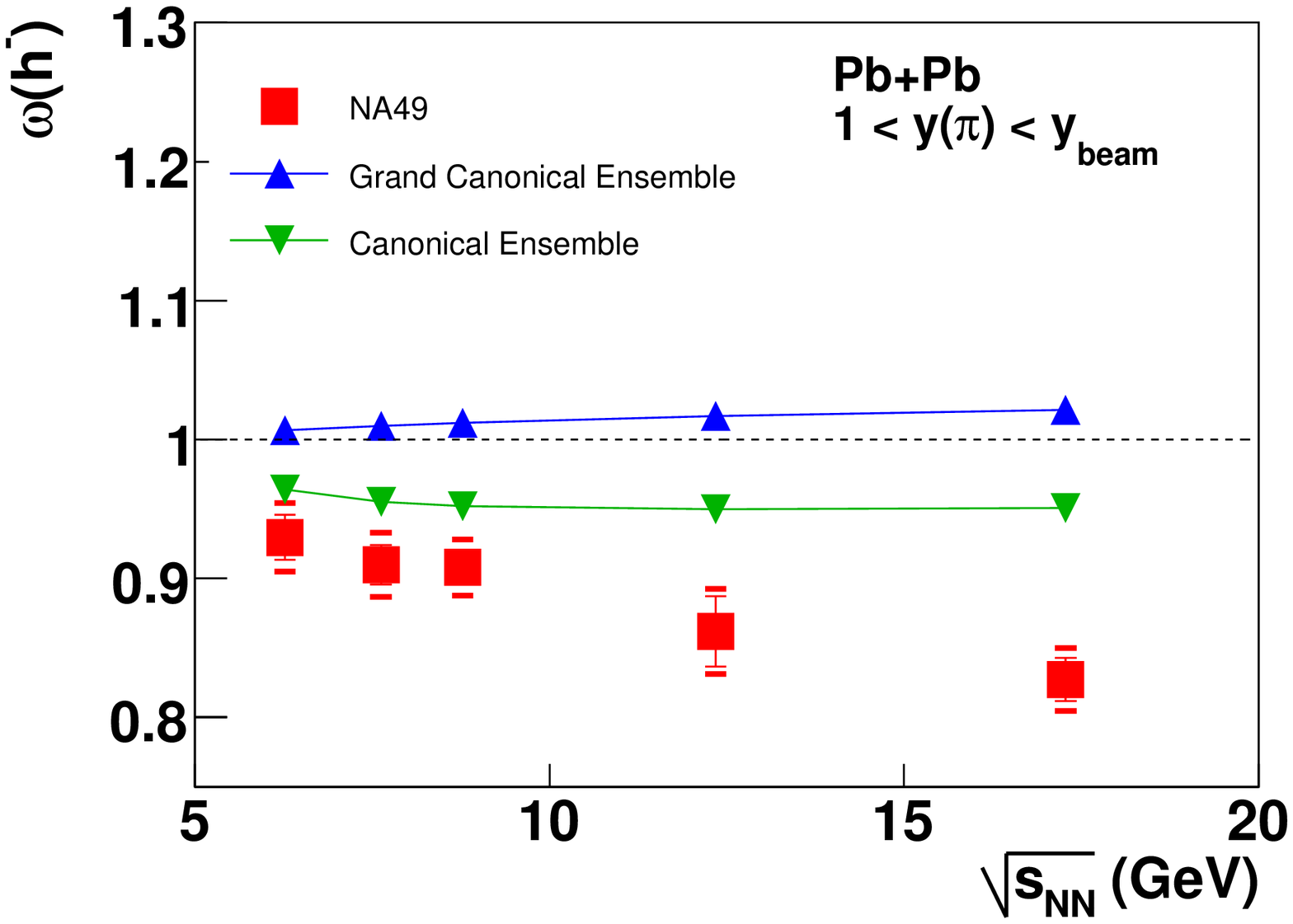}
\end{minipage}
\begin{minipage}{7cm}
\includegraphics[width=.99\textwidth]{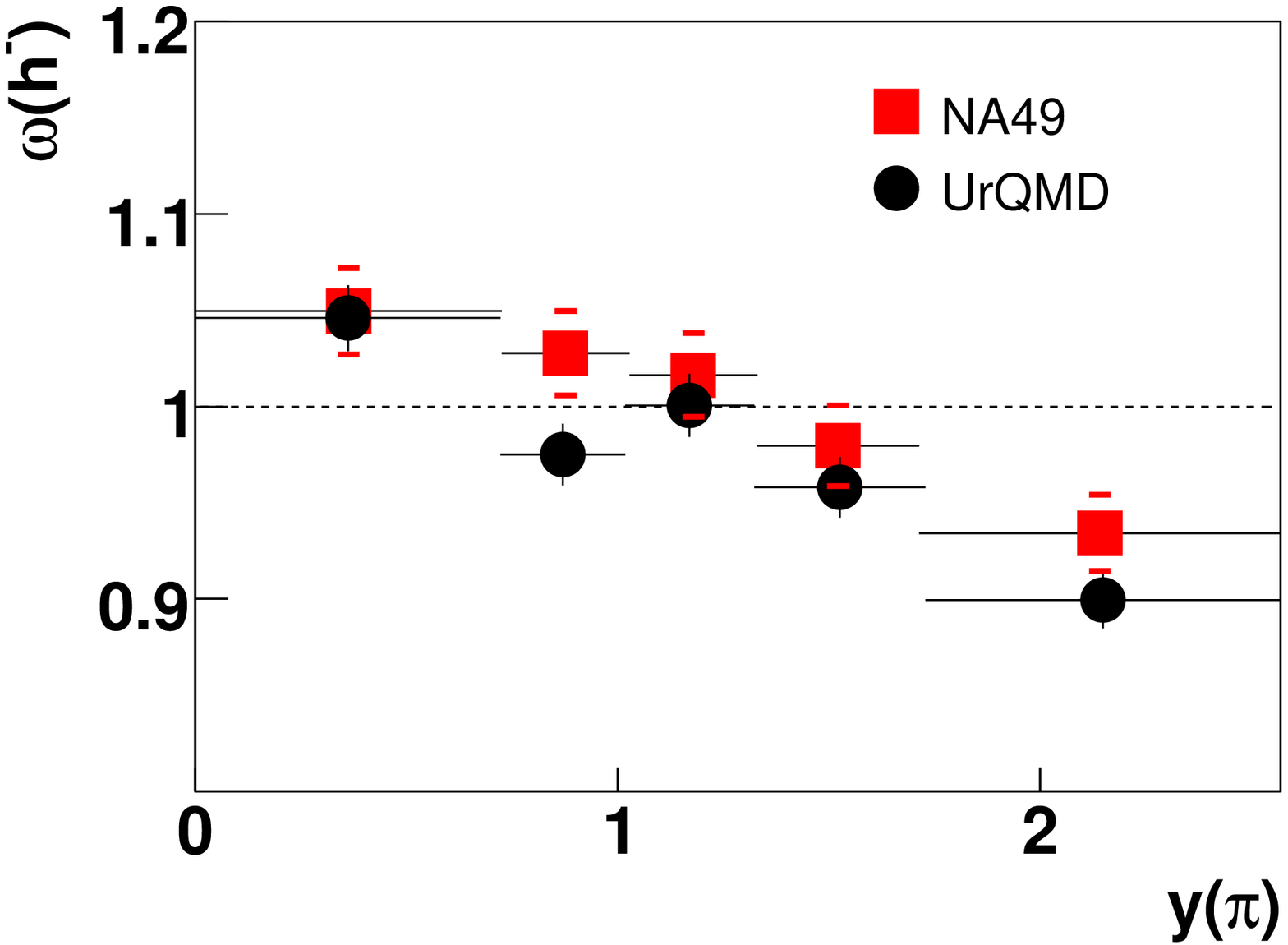}
\end{minipage}

\begin{center}
\begin{minipage}{11.8cm}
\begin{center}
\parbox[t]{5.2cm}{
\footnotesize{{\bf Figure 3:} (Color online)
\label{ed_statmod}Energy dependence of $\omega(h^-)$ in Pb+Pb
collisions in the rapidity interval $1<y<y_{beam}$ compared to the
predictions of a hadron-gas models~\cite{Begun:2006uu}. The lines
interpolate between the points calculated within the
models.}}\hfill
\parbox[t]{5.2cm}{
\footnotesize{{\bf Figure 4:} (Color online) \label{ypt_dep}Rapidity dependence of $\omega(h^-)$ at $80A$ GeV compared to UrQMD calculations.
}}
\end{center}
\end{minipage}
\end{center}
\end{figure}
In this section we compare the NA49 data on multiplicity
fluctuations to the predictions of hadron-gas
models~\cite{Begun:2006uu} as well as the string hadronic model
UrQMD~\cite{UrQMD}. Note, that the energy dependence of
multiplicity fluctuations is also predicted within the HSD
model~\cite{Konchakovski:2007ss}, but calculations for the NA49
acceptance are not published.

The predictions of the hadron-gas model in its grand-canonical and canonical formulation can be interpolated
between full and vanishing acceptance to the experimental acceptance~\cite{Begun:2006uu}, because in these ensembles the particles are essentially uncorrelated in momentum space and the particle momentum distribution is independent of multiplicity.
Figure~3 shows that the measured fluctuations disagree with the predictions
of the canonical and grand-canonical model in the forward acceptance. For the micro-canonical model no predictions
in the limited acceptance are available yet, but smaller fluctuations are expected.
The suppression of fluctuations in the experimental data with respect to the grand-canonical model can be
explained by charge and energy-momentum conservation~\cite{Begun:2006uu}.

The rapidity dependence of $\omega$ is shown in figure~4. The width of the $y$
bins is chosen in such a way that the mean multiplicity in
each bin is the same. A decrease of $\omega$ with increasing
rapidity is observed, which is reproduced by the UrQMD
model~\cite{Lungwitz:2007uc}. In terms of a hadron gas model this
can be understood as a result of energy and momentum
conservation~\cite{Hauer:2007im}.

\section{Conclusions}

Our main results from the study of central Pb+Pb collisions are as follows:
\begin{itemize}

\item No structure related to the critical point were observed in
the energy dependence of multiplicity fluctuations. Statistical
models show that conservation laws have an important effect on
$\omega$. The UrQMD model reproduces the trends observed in the
data.

\item There is no significant energy dependence of
transverse-momentum fluctuations at SPS energies. Thus the energy
dependence of $p_{T}$ fluctuations shows no evidence of the
critical point.

\item Net-charge fluctuations are independent of energy and
detector acceptance. Their values are close to zero, much above
the negative values expected for a QGP.

\item Dynamical kaon/pion fluctuations increase towards lower beam
energy. Fluctuations are significantly enhanced over hadronic
cascade models like UrQMD.

\end{itemize}

\vspace*{0.4cm}

{\bfseries Acknowledgements:} This work was supported by the US Department of Energy
Grant DE-FG03-97ER41020/A000,
the Bundesministerium fur Bildung und Forschung, Germany,
the Virtual Institute VI-146 of Helmholtz Gemeinschaft, Germany,
the Polish State Committee for Scientific Research (1 P03B 006 30, 1 P03B 097 29, 1 PO3B 121 29, 1 P03B 127 30),
the Hungarian Scientific Research Fund (OTKA 68506),
the Polish-German Foundation, the Korea Science \& Engineering Foundation (R01-2005-000-10334-0),
the Bulgarian National Science Fund (Ph-09/05) and the Croatian Ministry of Science, Education and Sport (Project 098-0982887-2878).

\end{document}